\documentclass[aps,prd,preprint,superscriptaddress,showkeys]{revtex4-1}

\usepackage{amsmath,amssymb,graphicx,subfigure}

\newcommand{\rmd}{\mathrm{d}}
\newcommand{\rme}{\mathrm{e}}

\begin{document}


\title{Naked singularity and black hole formation in self-similar \\ Einstein-scalar fields with exponential potentials}


\author{Xuefeng Zhang}
\email[]{zxf@unm.edu}
\affiliation{Department of Astronomy, Department of Physics, Beijing Normal University, Beijing 100875, China}

\author{Xinliang An}
\email[Joint first author, ]{xan@math.rutgers.edu}
\affiliation{Department of Mathematics, Rutgers University, Piscataway, NJ 08854-8019, USA}



\date{\today}

\begin{abstract}
  Motivated by cosmic censorship in general relativity and string theory, we extend Christodoulou's celebrated examples of naked singularity formation in the Einstein-massless scalar field system to include a positive or negative scalar potential of exponential forms, i.e., $V(\phi)=\pm\exp(2\phi/\kappa)$ with a parameter $\kappa$. Under spherical symmetry and a continuously self-similar ansatz depending on $\kappa$, we derive a 3-dimensional autonomous system of first-order ordinary differential equations, which incorporates the equations for massless scalar fields as a special case. Local behavior of the phase space is studied analytically with global solutions constructed numerically. Within the 3-dimensional solution manifold, we observe, for the negative potentials, naked singularity formation from nonsingular initial data for $\kappa^2<1$. Meanwhile, transitions between solutions containing naked singularities and black holes are also identified. However, when the potential is taken positive, numerical evolutions result in formation of black holes, but not naked singularities.
\end{abstract}


\maketitle

\section{Introduction}

A (globally) naked singularity in general relativity is a singularity visible to distant observers. The weak cosmic censorship conjecture \cite{Penrose69} postulates that generically naked singularities do not occur from gravitational collapse of physically reasonable matter. In other words, singularities should always be preceded by a trapped region without future light cones expanding to infinity. During the past few decades, only limited progress has been made toward a general proof or disproof of the conjecture \cite{Wald97,Penrose99,Rendall05}. From 1980s to late 90s, Christodoulou investigated global dynamics of the Einstein-scalar field equations with a vanishing potential in a series of papers. Remarkably, he established the first example of naked singularities that can evolve from regular asymptotically flat initial data in a suitable matter field \cite{Christo94}. Later on, he also proved that such occurrences are unstable and non-generic \cite{Christo99}, hence not violating weak cosmic censorship. By construction, his family of exact spacetimes in \cite{Christo94} is continuously self-similar and has \emph{two} real parameters $k$ ($\kappa$ in our notation) and $a$. Detailed analysis was given to the moment when the curvature singularity first appears and the self-similar coordinate becomes singular as well. A naked singularity arises provided $0<k^2<1/3$. The asymptotically flat initial data sets are obtained by truncating the self-similar ones. Relevant to this work, Brady examined the same model using numerical simulation \cite{Brady95} with more emphasis on critical behavior. Bedjaoui et al. further applied the self-similar construction to the Brans-Dicke theory \cite{Bedjaoui10} and showed that naked singularities can also emerge. In parallel to Christodoulou's work on massless scalar fields, Choptuik discovered a different type of naked singularities, admitting discrete self-similarity instead, in critical phenomena through fine-tunings of initial configurations (\cite{Choptuik93}, see also \cite{Martin03}). Since then, critical phenomena have been reported in many other systems (see the recent review \cite{Gundlach07}), for instance, in the Einstein-scalar field system with inclusion of nontrivial self-interaction potentials (quadratic or double-well forms \cite{Brady97,Hawley00,Honda02}). For more discussion on self-similarity in general relativity, see, e.g., \cite{Carr99,Carr05} and references therein.

In the context of anti-de Sitter/conformal field theory correspondence (AdS/CFT), there has been a series of discussions on possible violation of cosmic censorship in asymptotically AdS spacetimes \cite{Hertog04,Hertog04a,Garfinkle04,Gutperle04,Frolov04,Dafermos05,Langfelder05}. Most attention has focused on the Einstein-scalar field system with a self-interaction scalar potential $V(\phi)$. Unfortunately, for non-vanishing potentials (possibly negative, bounded or unbounded from below), the question of weak cosmic censorship in spherical symmetry remains open \cite{Dafermos05}. Examples of naked singularity formation analogous to Christodoulou's appear to be lacking.

In this paper, we aim to show that Christodoulou's self-similar model of naked singularities can be extended to include exponential potentials. In current literature, exponential potentials (Liouville type) often arise from scale-invariant supergravity theories as well as the Kaluza-Klein dimensional reduction on curved extra dimensions \cite{Townsend01,Duff86,Overduin97,Halliwell87}. They have also been intensively studied in scalar field cosmology \cite{Halliwell87,Coley03}. Most importantly for our purposes, the exponential forms are compatible with the assumption of continuous self-similarity, which makes our attempt viable in the first place.

In presenting our model, we will follow Brady's precedent of \cite{Brady95} with certain modifications (local compactification of the phase space) in order to exhibit the naked singularity solutions more prominently in numerical plots. When it comes to analysis of the equations and spacetimes, Christodoulou's paradigm is respected in our treatment. In particular, we can embed his 2-dimensional solution manifold into our 3-dimensional manifold, allowing us to draw analogy along the way.

The paper is structured as follows. In the next section, we introduce the self-gravitating scalar model and reduce the Einstein equations to an autonomous system of evolutionary equations. In Sec. \ref{sec:local}, local studies are performed on critical and singular points of the phase space. Then in Sec. \ref{sec:global}, we piece local pictures together and solve the equations by numerics in two parameter regimes $\kappa>1$ and $0<\kappa<1$, respectively. Afterwards, Sec. \ref{sec:struc} is devoted to analysing the spacetime structure of the naked singularity solutions, and concluding remarks are made in Sec. \ref{sec:conclusion}. For interested readers in Christodoulou's paper \cite{Christo94}, the Appendix \ref{app:Christo} contains our autonomous equations using his original notation.

\section{The field equations}

We follow mostly Brady's notation in \cite{Brady95} for an easier comparison. To start with, the Lagrangian density is chosen as
\begin{eqnarray}
 \mathcal{L} &=& \sqrt{-g} \left(\frac{R}{4} - \frac{1}{2}g^{\mu\nu}\nabla_\mu \phi \nabla_\nu \phi - V(\phi)\right), \label{Lag} \\
 V(\phi) &=& V_0 \exp\!\left(\frac{2\phi}{\kappa}\right), \qquad \kappa\neq 0,
 \qquad V_0=\pm 1\ \text{or } 0,
 \label{Vphi}
\end{eqnarray}
where $\phi$ is a real scalar field and $V(\phi)$ its self-interaction potential with a parameter $\kappa$. By a translation of $\phi$, one can normalize the multiplicative factor $V_0\neq 0$ to $\pm 1$. The energy-momentum tensor is given by
\begin{equation} \label{T}
 T_{\mu\nu} = \nabla_\mu \phi \nabla_\nu \phi - \left[\frac{1}{2}\left(\nabla_\alpha \phi \nabla^\alpha \phi\right) + V(\phi)\right] g_{\mu\nu}.
\end{equation}
The Einstein equations ($G_{\mu\nu}=2T_{\mu\nu}$) and the generalized Klein-Gordon equation are
\begin{eqnarray}
 R_{\mu\nu} &=& 2\big(\nabla_\mu \phi \nabla_\nu \phi + V(\phi)g_{\mu\nu}\big), \\
 \nabla_\alpha \nabla^\alpha \phi &=& V'(\phi).
\end{eqnarray}
Using retarded Bondi coordinates in spherical symmetry, we consider the following metric ($g$, $\bar{g}$ real-valued):
\begin{equation} \label{metric}
 \rmd s^2 = -g(u,r) \bar{g}(u,r)\,\rmd u^2 - 2 g(u,r)\,\rmd u\rmd r + r^2 \rmd\Omega^2,
\end{equation}
with $r$ the area radius and $\rmd\Omega^2$ the standard metric of a unit 2-sphere.
Then the field equations and the wave equation for $\phi=\phi(u,r)$ can be written as
\begin{eqnarray}
 (\ln g)_{,r} &=& r (\phi_{,r})^2, \label{feq1} \\
 (r \bar g)_{,r} &=& \left(1 - 2r^2 V(\phi)\right) g, \label{feq2} \\
 g(\bar{g}/g)_{,u} &=& 2r \left[(\phi_{,u})^2 - \bar{g} \phi_{,u}\phi_{,r}\right], \label{feq3} \\
 r^{-1} (r^2 \bar{g}\phi_{,r})_{,r} &=& 2\phi_{,u} + 2r\phi_{,ru} + r g V'(\phi). \label{feq4}
\end{eqnarray}
For the center $r=0$ to be regular (finite Ricci scalar), we take
\begin{equation} \label{BV}
 g(u,0) = \bar{g}(u,0) = 1,
\end{equation}
which also fixes the null coordinate $u$ as the proper time of an observer at the center. Additionally, the Misner-Sharp (Hawking) mass is defined by
\begin{equation} \label{mass}
 1-\frac{2m(u,r)}{r} = g^{\mu\nu} r_{,\mu} r_{,\nu} = \frac{\bar g}{g}.
\end{equation}

To study self-similar solutions, we notice that including an exponential potential as given in (\ref{Vphi}) does not disrupt the existence of a homothetic Killing vector ($L_\xi g_{\mu\nu} = 2g_{\mu\nu}$)
\begin{equation}
 \xi = u \partial_u + r \partial_r
\end{equation}
for the metric (\ref{metric}). This allows us to adopt the same self-similar ansatz for the massless scalar field (see \cite{Christo94} and the Appendix A of \cite{Brady95} for detailed derivation):
\begin{eqnarray}
 \rmd s^2 &=& -g(x) \bar{g}(x)\, \rmd u^2 - 2 g(x)\, \rmd u\rmd r + r^2 \rmd\Omega^2, \label{metricx} \\
 \phi &=& \bar{h}(x) - \kappa \ln(-u), \qquad x=-\frac{r}{u}, \qquad u<0,
\end{eqnarray}
with $\bar{h}$ a real-valued unknown to be determined. Apparently, the self-similar coordinate $x$ becomes singular at $u=0$, and we are primarily concerned with the past region $u<0$ as $u$ increases toward the future. Moreover, note that the constant $\kappa$ in the ansatz also serves as a parameter in the scalar potential (\ref{Vphi}), and hence is fixed by the Lagrangian if $V_0\neq 0$. One can check that this particular arrangement is necessary in order to reduce the field equations to ordinary differential equations (ODEs).

Derived from (\ref{feq1}-\ref{feq4}), the resulting equations for the four unknowns---$g(x)$, $\bar{g}(x)$, $\bar{h}(x)$, and $\gamma(x)$ (defined in (\ref{fode1}))---consist of four first-order ODEs ($\ '=\rmd/\rmd x$) plus one algebraic equation:
\begin{eqnarray}
 x \bar{h}' &=& \gamma, \label{fode1} \\
 x g' &=& g\gamma^2, \\
 x\bar{g}' + \bar{g} &=& \big(1 - 2V_0 x^2 \rme^{2\bar{h}/\kappa}\big)\, g, \\
 (\bar{g} - 2x)\, (x\gamma') &=& 2\kappa x - (g - 2x)\, \gamma + 2V_0 x^2 \rme^{2\bar{h}/\kappa} g (\gamma + \kappa^{-1}), \label{fode4} \\
 g - \bar{g} &=& 2\kappa^2 x - (\bar{g} - 2x)(\gamma^2 + 2\kappa\gamma) + 2V_0 x^2 \rme^{2\bar{h}/\kappa} g.
\end{eqnarray}
One may check that the derivative of the last equation, which derives from (\ref{feq3}), is a combination of (\ref{fode1}-\ref{fode4}). If $V_0=0$, the equations simplify to the special case for massless scalar fields (see \cite{Brady95}, (2.9-2.13)).

To convert the above equations into an autonomous system, we follow Brady's method and introduce new unknown variables:
\begin{equation} \label{wyzs}
 w(s) = x^2 \rme^{2\bar{h}/\kappa} \geq 0, \qquad y(s) = \frac{\bar g}{g},
 \qquad z(s) = \frac{x}{\bar g}, \qquad x = \rme^s \geq 0.
\end{equation}
Thus we obtain ($\ \dot{} = \rmd/\rmd s = x\rmd/\rmd x$; cf. \cite{Brady95}, (2.18-2.21))
\begin{eqnarray}
 \dot{w} &=& 2w (\gamma + \kappa)/\kappa, \label{dew} \\
 \dot{y} &=& 1 - y\,(1+\gamma^2) - 2V_0 w, \label{dey} \\
 \dot{z} &=& z\left[2 - \left(1-2V_0 w \right)y^{-1}\right], \label{dez} \\
 (1-2z)\dot{\gamma} &=& 2\kappa z - \gamma(y^{-1}-2z) + 2V_0 w y^{-1} (\gamma+\kappa^{-1}), \label{degamma} \\
 (1-2z)(\gamma+\kappa)^2 &=& 1 + \kappa^2 - (1 - 2V_0 w) y^{-1}, \qquad V_0=\pm 1\ \text{or } 0.
 \label{eqa}
\end{eqnarray}
It can be verified that (\ref{dey}-\ref{eqa}) imply (\ref{dew}). Hence using (\ref{eqa}), we can remove the term $V_0 w$ in (\ref{dey}-\ref{degamma}) and reduce the problem to a 3-dimensional dynamical system:
\begin{eqnarray}
 \dot{y} &=& 2(\gamma+\kappa)\left[(\gamma+\kappa)z-\gamma\right] y, \label{dey3} \\
 \dot{z} &=& -2(\gamma+\kappa)^2 z^2 + \left[(\gamma+\kappa)^2+1-\kappa^2\right] z, \label{dez3} \\
 (1-2z)\dot{\gamma} &=& (1-2z)\left[\gamma^3 + \left(\frac{1}{\kappa}+2\kappa\right)\gamma^2\right]
 + (1-2z-2\kappa^2 z)\gamma - \frac{1}{\kappa}\left(1 - \frac{1}{y}\right). \label{degamma3}
\end{eqnarray}
Given these equations, $w$ as determined by (\ref{eqa}) satisfies (\ref{dew}). Other ways to bring about autonomous systems are also possible. For instance, see the Appendix \ref{app:Christo} for comparison with Christodoulou's treatment and notation.

The system (\ref{dey3}-\ref{degamma3}), as derived from (\ref{dew}-\ref{eqa}), automatically includes Brady's special case for massless scalars, i.e., (\ref{dew}-\ref{eqa}) with $w=0$ (see also \cite{Brady95}, (2.18-2.20)). One can retrieve the latter from (\ref{dey3}-\ref{degamma3}) by imposing the algebraic constraint (\ref{eqa}) with $V_0 w=0$.

To facilitate analysis of the autonomous system, we further introduce
\begin{equation} \label{zzeta}
 z = \frac{\zeta}{2(1-\zeta)}.
\end{equation}
This substitution is intended to bring $z=+\infty$ to a finite region (compactification), i.e., $\zeta=1$, as we will see that naked singularity solutions tend to $z=+\infty$ as $s\rightarrow +\infty$. In terms of $y$, $\zeta$, and $\gamma$, the system (\ref{dey3}-\ref{degamma3}) takes the form
\begin{eqnarray}
 \dot{y} &=& \frac{y(\gamma+\kappa)(3\gamma\zeta - 2\gamma + \kappa\zeta)}{1-\zeta}, \label{ode1} \\
 \dot{\zeta} &=& -\left[2(\gamma+\kappa)^2 + 1 - \kappa^2\right] \zeta^2 + \left[(\gamma+\kappa)^2+1-\kappa^2\right] \zeta, \label{ode2} \\
 \dot{\gamma} &=& \gamma^3 + \left(\kappa^{-1} + 2\kappa\right)\gamma^2
 + \left(1 - \frac{\kappa^2 \zeta}{1-2\zeta}\right)\!\gamma - \frac{1-\zeta}{\kappa(1-2\zeta)}\left(1 - \frac{1}{y}\right), \label{ode3}
\end{eqnarray}
and the constraint for massless scalar fields ((\ref{eqa}) with $V_0 w=0$) reads
\begin{equation} \label{eqaml}
 (\gamma+\kappa)^2 \frac{1-2\zeta}{1-\zeta} = 1 + \kappa^2 - \frac{1}{y}.
\end{equation}
Furthermore, the initial conditions imposed by regularity at the center (cf. (\ref{BV})) require
\begin{equation} \label{IV}
 w\rightarrow 0, \qquad y\rightarrow 1, \qquad \zeta\rightarrow 0, \qquad \gamma\rightarrow 0,
 \qquad \textrm{as} \qquad s\rightarrow -\infty \ \ (x\rightarrow 0).
\end{equation}
The objective is to solve the evolutionary equations (\ref{ode1}-\ref{ode3}) from $s=-\infty\ (u<0,r=0)$ to possibly $s=+\infty\ (u=0,r>0)$. Among solutions satisfying (\ref{IV}), black hole formation is indicated by integral curves reaching the plane $y=0$, where it locates an apparent horizon (cf. (\ref{mass})). Hence for naked singularities, we need to identify solutions that can avoid contacting $y=0$ for the entire range of $s$. For this purpose, solutions terminating at a finite critical or singular point as $s\rightarrow +\infty$ are reasonable candidates. Moreover, requiring that the mass function be non-negative implies $y\leq 1$, which can be used to narrow the search.

In addition, the system (\ref{ode1}-\ref{ode3}) is invariant under the mapping $\kappa\rightarrow -\kappa$, $\gamma\rightarrow -\gamma$ with $y$ and $\zeta$ unchanged. Without loss of generality, we only consider $\kappa>0$ henceforward.

\section{Local behaviors} \label{sec:local}

Before we investigate global solutions, it is important to understand local phase space near critical and singular points (see Table \ref{table} for a quick summary). At these key locations, we can determine properties such as stability, dimensions of stable/unstable manifolds, uniqueness of solutions, etc., to help us build a fuller picture. Because the phase space for the massless scalar field is contained as a surface (signified by the constraint (\ref{eqaml})) in our 3-dimensional phase space, we expect that many discussions on this surface \cite{Christo94} can be transferable to the whole space.

\subsection{The initial point $s = -\infty$} \label{subsec:IP}

The initial point
\begin{equation}
 \mathcal{O}: \ \ y=1, \ \ \zeta=0, \ \ \gamma=0,
\end{equation}
is also a critical point of the vector field defined by (\ref{ode1}-\ref{ode3}). Therefore, we can determine the behavior of solutions in a neighborhood of $\mathcal{O}$ by the standard linearization method. With
\begin{equation}
 y = 1 + x_1, \qquad \zeta = x_2, \qquad \gamma = x_3,
\end{equation}
the Taylor series expansion in $x_{1,2,3}$ gives rise to a linearized version of (\ref{ode1}-\ref{ode3}):
\begin{equation} \label{lineq1}
 \dot{\mathbf{x}} = A\,\mathbf{x}, \qquad \mathbf{x} = (x_1,x_2,x_3)^T,
 \qquad A =
 \begin{pmatrix}
 0 & \kappa^2 & -2\kappa \\
 0 & 1 & 0 \\
 -\kappa^{-1} & 0 & 1
 \end{pmatrix}
 .
\end{equation}
The Jacobian matrix $A$ has eigenvalues
\begin{equation} \label{eigen_O}
 \lambda_1 = -1, \qquad \lambda_2 = 1, \qquad \lambda_3 = 2.
\end{equation}
Thus $\mathcal{O}$ is a saddle point. The general solution of (\ref{lineq1}) is given by
\begin{equation}
 x_1 = c_1 \rme^{\lambda_1 s} - c_3 \rme^{\lambda_3 s}, \qquad x_2 = c_2 \rme^{\lambda_2 s},
 \qquad x_3 = \frac{c_1 \rme^{\lambda_1 s} + c_2 \kappa^2 \rme^{\lambda_2 s} + 2c_3 \rme^{\lambda_3 s}}{2\kappa},
\end{equation}
with three integration constants $c_{1,2,3}$. Now we recall the conditions (\ref{BV}), (\ref{wyzs}), and (\ref{zzeta}). They together imply
\begin{equation} \label{IV1}
 \lim_{s\rightarrow-\infty} \zeta\, \rme^{-s} = 2,
\end{equation}
and hence set $c_2=2$. Consequently, solutions subject to the conditions (\ref{IV}) and (\ref{IV1}) take the form
\begin{equation} \label{seriesL}
 y = 1 - c_3 \rme^{2s} + O(x_i x_j), \qquad \zeta = 2\rme^s + O(x_i x_j),
 \qquad \gamma = \kappa \rme^s + \frac{c_3}{\kappa}\rme^{2s} + O(x_i x_j),
\end{equation}
which has one free parameter $c_3$ associated with $\lambda_3$ and $i,j=1,2,3$. Particularly, the massless solution, which satisfies (\ref{eqaml}), occurs uniquely at $c_3=7\kappa^2/3$. The presence of a free parameter can also be confirmed by the Taylor series solution at $\mathcal{O}$ in terms of $x$. More specifically, we have
\begin{eqnarray}
 y &=& 1 - c_T x^2 - 2c_T x^3 + O(x^4), \nonumber \\
 \zeta &=& 2x - 4x^2 + (2c_T+8-\kappa^2)x^3 + O(x^4), \label{seriesT} \\
 \gamma &=& \kappa x + \left(\frac{c_T}{\kappa}+\kappa\right)\! x^2
 + \left[\frac{3c_T}{\kappa} + (c_T+1)\kappa - \frac{\kappa^3}{2}\right]\! x^3 + O(x^4), \nonumber
\end{eqnarray}
with $c_T=\kappa^2/3$ for the unique massless solution. These series solutions make up our first step toward global solutions. They will also be useful for generating approximate initial values near $\mathcal{O}$ for numerical integration.

Compared to Christodoulou's massless case, the interior solution is no longer unique. Instead, we have a 1-parameter family of integral curves emitting from $\mathcal{O}$. This enlarged solution manifold is due to the addition of the third eigenvalue $\lambda_3$ ($\lambda_{1,2}$ identical with the massless case \cite{Christo94}), which will lead to richer behavior.

\begin{figure}[htbp]
  \centering
  \mbox{
  \subfigure{\includegraphics[width=0.48\textwidth]{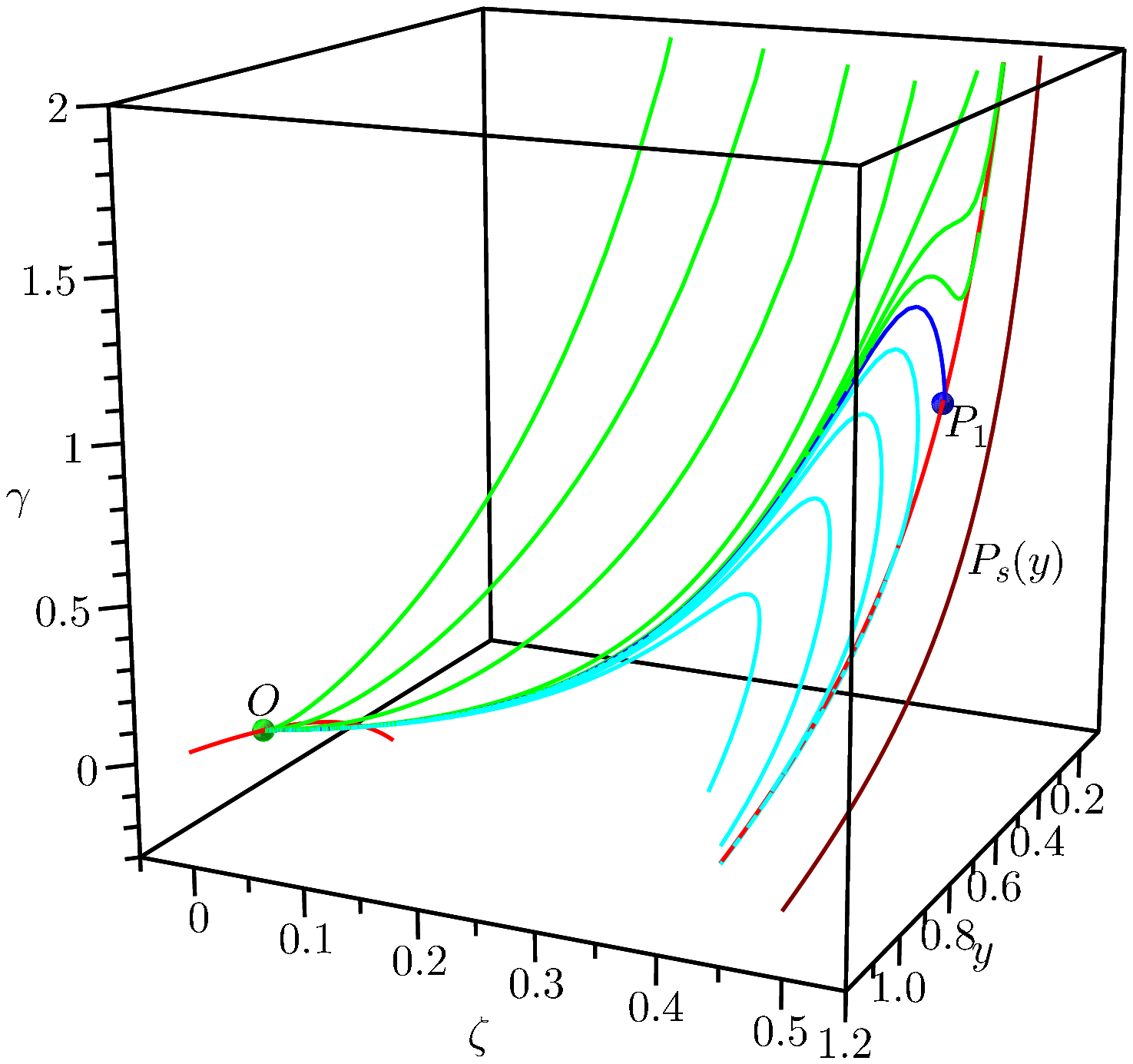}}
  \subfigure{\includegraphics[width=0.48\textwidth]{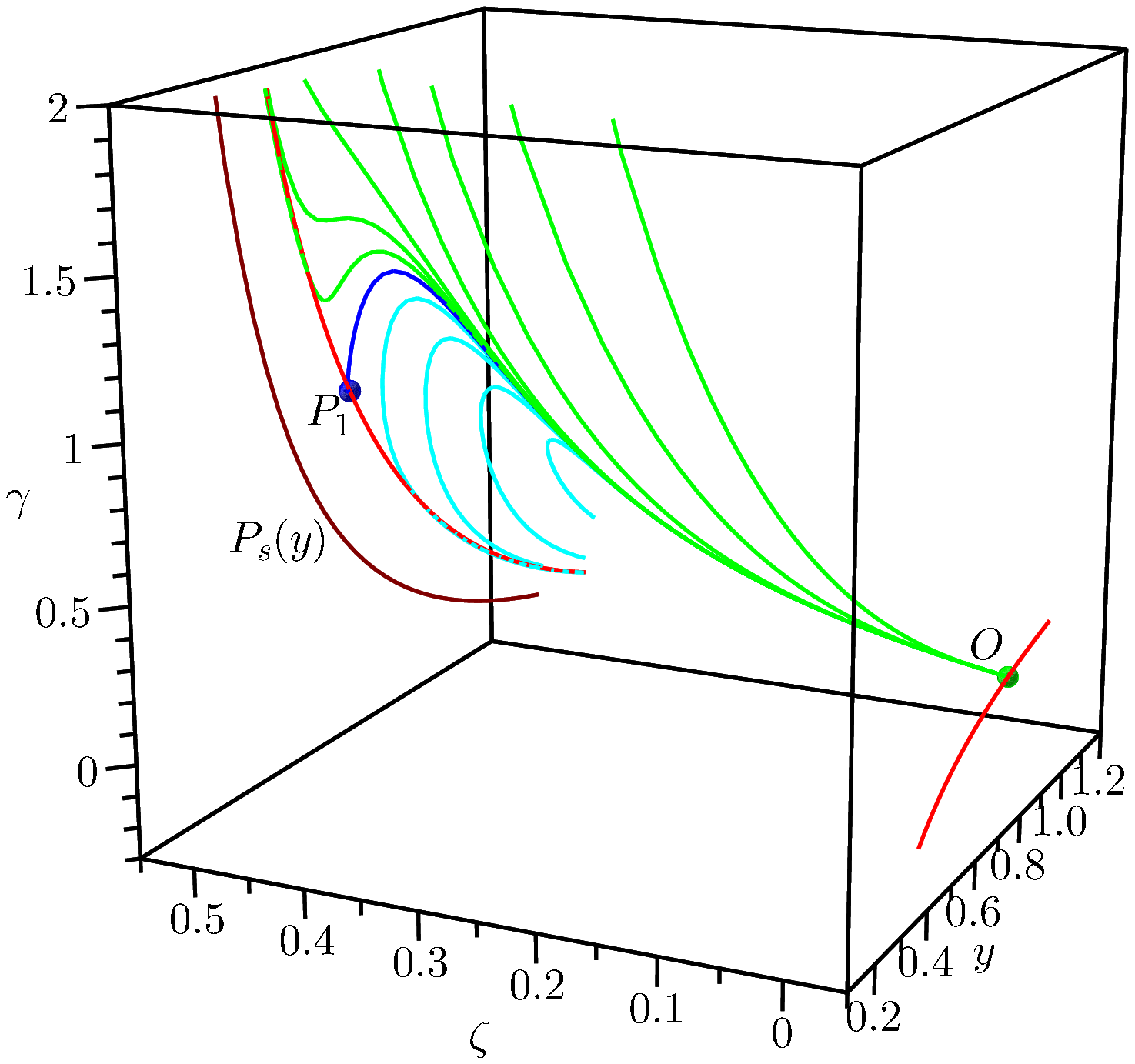}}
  }
  \caption{Numerical integration with $\kappa=1.2$. The right plot shows the back view of the left. Green ($c_T>\kappa^2/3$, $V_0=1$, cf. (\ref{V0_cT})), blue, and cyan ($c_T<\kappa^2/3$, $V_0=-1$) curves emanating from $\mathcal{O}$ are members of the 1-parameter family of solutions of the system (\ref{ode1}-\ref{ode3}) subject to (\ref{IV}) and (\ref{IV1}). The negatively unstable manifold of $\mathcal{O}$ and the unstable manifold of $\mathcal{P}_1$ are marked by red curves. The blue curve ($c_T=\kappa^2/3$, $V_0=0$) ending at $\mathcal{P}_1$ delineates the unique massless solution. The maroon curve represents the set of singular points $\mathcal{P}_s(y)$ with $\zeta=1/2$. }
  \label{fig1}
\end{figure}

So far, we have not specified the sign of the scalar potential since the factor $V_0$ does not appear in (\ref{ode1}-\ref{ode3}). To determine the sign, we plug the series solution (\ref{seriesT}) into (\ref{eqa}) and obtain
\begin{equation}
 0 \leq \rme^{2\bar{h}/\kappa} = \frac{w}{x^2} = \frac{3c_T-\kappa^2}{2V_0} (1+2x) + O(x^2).
\end{equation}
For a real scalar field, the above quantity must be kept non-negative, which then sets
\begin{equation} \label{V0_cT}
 V_0 = \left\{
 \begin{array}{rl}
 -1 &\ \text{if } c_T < \kappa^2/3 \ \ (c_3 < 7\kappa^2/3), \\
 +1 &\ \text{if } c_T > \kappa^2/3 \ \ (c_3 > 7\kappa^2/3).
 \end{array} \right.
\end{equation}
Likewise, using (\ref{seriesL}), we have an equivalent sign choice in terms of $c_3$ given in the parentheses above. Therefore, based on the sign of $V_0$, the 2-dimensional interior solution manifold is divided into two regions by the unique massless solution ($c_T=\kappa^2/3$, $V_0=0$) (cf. Fig. \ref{fig1}).

\subsection{Critical points $s=+\infty$} \label{sec:critpt}

Besides the initial point, the system possesses three more critical points that may serve as ending points of solution curves. We can treat them by linearization as well.

For the first critical point
\begin{equation}
 \mathcal{P}_1: \ \ y = \frac{1}{2},\ \ \zeta = \frac{2}{3+\kappa},\ \ \gamma = 1,
\end{equation}
the matrix $A$ reads
\begin{equation}
 A =
 \begin{pmatrix}
 0 & \frac{1}{2}(3+\kappa)^2 & -\kappa \\
 0 & -2(1+\kappa) & -\frac{4(1-\kappa)(1+\kappa)}{(3+\kappa)^2} \\
 \frac{4(1+\kappa)}{\kappa(1-\kappa)} & \frac{(1+\kappa+\kappa^2)(\kappa+3)^2}{\kappa(1-\kappa)}
 & \frac{2(1+\kappa-\kappa^3)}{\kappa(1-\kappa)}
 \end{pmatrix},
\end{equation}
with eigenvalues
\begin{equation} \label{eigen_P1}
 \lambda_1 = \frac{\kappa-\sqrt{4-3\kappa^2}}{1-\kappa},
 \qquad \lambda_2 = \frac{\kappa+\sqrt{4-3\kappa^2}}{1-\kappa},
 \qquad \lambda_3 = \frac{2(1+\kappa)}{\kappa}.
\end{equation}
Particularly, we have $\mathrm{Re}\lambda_{1,2}<0$ and $\lambda_3>0$ when $\kappa>1$, and $\lambda_1<0$, $\lambda_2>0$ and $\lambda_3>0$ when $0<\kappa<1$. Thus $\mathcal{P}_1$ is a saddle point. For the second critical point
\begin{equation}
 \mathcal{P}_{-1}: \ \ y = \frac{1}{2},\ \ \zeta = \frac{2}{3-\kappa},\ \ \gamma = -1,
\end{equation}
the matrix $A$ reads
\begin{equation}
 A =
 \begin{pmatrix}
 0 & \frac{1}{2}(3-\kappa)^2 & -\kappa \\
 0 & -2(1-\kappa) & \frac{4(1-\kappa)(1+\kappa)}{(3+\kappa)^2} \\
 \frac{4(1-\kappa)}{\kappa(1+\kappa)} & -\frac{(1-\kappa+\kappa^2)(\kappa-3)^2}{\kappa(1+\kappa)}
 & -\frac{2(1-\kappa+\kappa^3)}{\kappa(1+\kappa)}
 \end{pmatrix},
\end{equation}
with eigenvalues
\begin{equation} \label{eigen_P-1}
 \lambda_1 = \frac{-\kappa-\sqrt{4-3\kappa^2}}{1+\kappa},
 \qquad \lambda_2 = \frac{-\kappa+\sqrt{4-3\kappa^2}}{1+\kappa},
 \qquad \lambda_3 = -\frac{2(1-\kappa)}{\kappa}.
\end{equation}
Particularly, we have $\mathrm{Re}\lambda_{1,2}<0$ and $\lambda_3>0$ when $\kappa>1$, and $\lambda_1<0$, $\lambda_2>0$ and $\lambda_3<0$ when $0<\kappa<1$. Thus $\mathcal{P}_{-1}$ is also a saddle point.

Both of these points reside in the massless phase subspace, and they are directly inherited from Christodoulou's case for having the same locations as well as the first two eigenvalues $\lambda_{1,2}$. Hence we have denoted them in the same way as in \cite{Christo94}.

The third critical point does not have a massless counterpart:
\begin{equation}
 \mathcal{P}_{c}: \ \ y = \frac{1}{1-\kappa^4},\ \ \zeta = 0,\ \ \gamma = -\kappa.
\end{equation}
However, as we will see in later sections, it is not of much relevance to our discussion. Here we only list the matrix $A$ and its eigenvalues below without further ado:
\begin{equation}
 A =
 \begin{pmatrix}
 0 & 0 & \frac{2\kappa}{1-\kappa^4} \\
 0 & 1-\kappa^2 & 0 \\
 -\frac{(1-\kappa^4)^2}{\kappa} & 0 & -1-\kappa^2
 \end{pmatrix},
\end{equation}
\begin{equation}
 \lambda_{1,2} = \frac{-1-\kappa^2 \mp \sqrt{(1+\kappa^2)(9\kappa^2-7)}}{2},
 \qquad \lambda_3 = 1-\kappa^2.
\end{equation}

\subsection{Singular points $\zeta = 1$, $s=+\infty$} \label{sec:singlpt}

Besides the points $\mathcal{P}_{\pm 1}$ , Christodoulou also discussed a third critical point
\begin{equation} \label{P0}
 \mathcal{P}_0: \ \ y=\frac{1}{1+\kappa^2},\ \ \zeta = 1,\ \ \gamma = -\kappa,
\end{equation}
in the 2-dimensional phase space, which turned out to be very important since solutions representing naked singularities tend to this point as $s\rightarrow +\infty$. For our 3-dimensional problem, $\mathcal{P}_0$ expands into a straight line in the $y$-direction:
\begin{equation} \label{P0y0}
 \mathcal{P}_0(y_0): \ \ y=y_0\neq 0,\ \ \zeta = 1,\ \ \gamma = -\kappa.
\end{equation}
Points on the line are critical for the equations (\ref{ode2}-\ref{ode3}), but singular for (\ref{ode1}).

To probe into the local behavior near $\mathcal{P}_0(y_0)$, we continue using
\begin{equation}
 y = y_0 + x_1, \qquad \zeta = 1+x_2, \qquad \gamma = -\kappa + x_3,
\end{equation}
as in previous subsections and perform (Laurent/power) series expansions in terms of $x_{1,2,3}$. Only keeping terms up to the first order, we arrive at
\begin{eqnarray}
 \dot x_1 &=& 2 y_0 \kappa x_3 - y_0 \frac{x_3^2}{x_2}, \label{x1de} \\
 \dot x_2 &=& (\kappa^2-1) x_2, \label{x2de} \\
 \dot x_3 &=& \left(\frac{1-y_0}{y_0\kappa}+\kappa^3\right) x_2 - x_3. \label{x3de}
\end{eqnarray}
It is noted that a partial linearization succeeds in the $(\zeta,\gamma)$-plane since the linear equations (\ref{x2de}) and (\ref{x3de}) do not depend on $x_1$. The matrix $A$ for $x_{2,3}$ alone reads
\begin{equation}
 \begin{pmatrix}
 \kappa^2-1 & 0 \\
 \frac{1-y_0}{y_0\kappa}+\kappa^3 & -1
 \end{pmatrix}.
\end{equation}
The eigenvalues are
\begin{equation} \label{eigen_P0}
 \lambda_1 = -1, \qquad \lambda_2 = -(1-\kappa^2),
\end{equation}
which coincide with the massless case (cf. \cite{Christo94}, (3.16)). Solving the linear equations gives rise to
\begin{equation} \label{P0_zgamma}
 \zeta = 1 + c_2 \rme^{-(1-\kappa^2)s} + \cdots,
 \qquad \gamma = -\kappa + c_1 \rme^{-s} + c_2 \left(\frac{1-y_0}{y_0\kappa^3}+\kappa\right)\rme^{-(1-\kappa^2)s} + \cdots,
\end{equation}
with two parameters $c_{1,2}$. Fortunately, the troublesome term $x_3^2/x_2$ in (\ref{x1de}) is well-behaved under (\ref{P0_zgamma}), and we obtain formally
\begin{eqnarray}
 y = y_0 + \frac{c_2(1-y_0+y_0\kappa^4)(1-y_0-y_0\kappa^4)}{\kappa^6(1-\kappa^2) y_0} \rme^{-(1-\kappa^2)s}
 + \frac{2c_1(1-y_0)}{\kappa^3} \rme^{-s} \nonumber \\
 + \frac{y_0 c_1^2}{c_2(1+\kappa^2)} \rme^{-(1+\kappa^2)s} + \cdots, \label{P0_y}
\end{eqnarray}
where quadratic and higher order terms in $x_{1,2,3}$ are omitted. Hence for $\kappa^2<1$, the asymptotic expansions suggest that each $\mathcal{P}_0(y_0)$ acts like an attractor (except in the $y$-direction). This will be later confirmed by numerical calculation. Very much like the massless solutions, $\mathcal{P}_0(y_0)$ will serve as the endpoints of naked singularity solutions (see Sec. \ref{sec:struc}).

\subsection{Singular (connecting) points $\zeta = \frac{1}{2}$, $s=s_*$}

If a continuous integral curve from $\mathcal{O}$ were to reach a point $\mathcal{P}_0(y_0)$ with $\zeta=1$, it would have to cross the plane $\zeta=1/2$ at the following set of singular points:
\begin{equation}
 \mathcal{P}_s(y_*): \ \ y = y_*\neq 0,\ \ \zeta = \frac{1}{2},\ \ \gamma = \frac{1-y_*}{y_*\kappa^3},
\end{equation}
which is imposed by (\ref{ode3}) (multiplied by $1-2\zeta$ on both sides). Hence if an interior solution attains $\mathcal{P}_s(y_*)$ at some finite $s=s_*$, it is necessary to extend it past $\zeta=1/2$.

To tackle these singular points where the uniqueness of solutions fails, we follow Christodoulou's method and adopt a new independent variable $t$ such that
\begin{equation}
 \frac{\rmd s}{\rmd t} = -2(1-2\zeta).
\end{equation}
In terms of $t$, the singular point $\mathcal{P}_{s}(y_*)$ becomes a critical point of the transformed equations. Then perform linearization with
\begin{equation}
 y = y_* + x_1(t), \qquad \zeta = \frac{1}{2} + x_2(t), \qquad \gamma = \frac{1-y_*}{y_*\kappa^3} + x_3(t),
\end{equation}
and calculate the matrix $A$ as
\begin{equation}
 \begin{pmatrix}
 0 & \frac{4(-1+y_*+y_*\kappa^4)(1-y_*+y_*\kappa^4)}{y_*\kappa^6} & 0 \\
 0 & 1-\kappa^2 & 0 \\
 \frac{1}{y_*^2\kappa} & \frac{4(1-y_*)(1-y_*+y_*\kappa^4)(1-y_*+y_*\kappa^2+y_*\kappa^4)}{y_*^3\kappa^9}
 & \kappa^2
 \end{pmatrix},
\end{equation}
along with its eigenvalues
\begin{equation} \label{eigen_Ps}
 \lambda_1 = \kappa^2, \qquad \lambda_2 = 1-\kappa^2, \qquad \lambda_3 = 0,
\end{equation}
which are independent of $y_*$. The zero eigenvalue $\lambda_3$ reflects the fact that the set of $\mathcal{P}_s(y_*)$ forms a curve, while the other two eigenvalues are identical to the massless case (cf. \cite{Christo94}, p. 620). Therefore, if $\kappa^2<1$, both $\lambda_{1,2}$ are positive and one can construct a 1-parameter family of solutions which tend to $\mathcal{P}_s(y_*)$ as $t\rightarrow -\infty$ (or $s\rightarrow s_*$ from either side of $\zeta=1/2$) \cite{Christo94}. Consequently, a single interior solution curve from $\zeta<1/2$ may branch out at $\mathcal{P}_s(y_*)$ into a 2-dimensional solution manifold beyond $s_*$. We will further test this claim by numerical calculation.

As an extra comment, we mention that the massless phase subspace determined by (\ref{eqaml}) only intersect the curve $\mathcal{P}_s(y)$ at one point:
\begin{equation}
 \mathcal{P}_{sc}:\ \ y = \frac{1}{1+\kappa^2}, \ \ \zeta = \frac{1}{2}, \ \ \gamma = \frac{1}{\kappa}.
\end{equation}
which was closely examined by Christodoulou \cite{Christo94}. In Sec. \ref{case2}, we will compare continuations of the solutions from $\mathcal{P}_{sc}$ with those from other points on $\mathcal{P}_s(y)$.

\subsection{Apparent horizon $y=0$}

The apparent horizon is signaled by $y=0$, which also renders the equation (\ref{ode3}) singular. To capture the dominant behavior near $y=0$, we perform the Painlev\'{e} analysis \cite{Conte08} and identify the following Puiseux series expansions:
\begin{eqnarray}
 y &=& a_1\sqrt{s_\infty-s} - \left[\frac{a_1\kappa^2}{3a_2} + 4a_1a_2
 \pm \left(\frac{4\sqrt{2}a_1\kappa}{3} + \frac{\sqrt{2}(2a_1a_2-1)}{3a_2\kappa} \right)\right](s_\infty-s)
 + \cdots, \\
 \zeta &=& \frac{1}{2} + a_2\sqrt{s_\infty-s} + \cdots, \\
 \gamma &=& \pm \frac{1}{\sqrt{2} \sqrt{s_\infty-s}} - \frac{2\kappa}{3} - \frac{2a_1a_2-1}{6a_1a_2\kappa}
 \mp \frac{\sqrt{2}\kappa^2}{12a_2} + \cdots, \label{gammaAH}
\end{eqnarray}
which converges to
\begin{equation}
 \mathcal{P}_{AH}=\mathcal{P}_s(0):\ \ y = 0,\ \ \zeta = \frac{1}{2},\ \ \gamma = \pm \infty,
\end{equation}
as $s\rightarrow s_\infty-$. This series solution is general for having three free parameters $a_1$, $a_2$ and $s_\infty$. Thereby, one may regard the apparent horizon $\mathcal{P}_{AH}$ effectively as an \emph{attractor} in the phase space. The leading order terms of the expansions agree with Christodoulou's Proposition 3.2 in \cite{Christo94}.

Using (\ref{gammaAH}), we can estimate the scalar field near the apparent horizon as
\begin{equation}
 \bar{h}(x) = \int \frac{\gamma(x)}{x} \rmd x
 \simeq \frac{1}{\sqrt{2}} \int \frac{1}{x \sqrt{\ln x_{AH} - \ln x}} \rmd x
 = -\sqrt{2 \ln\!\left(\frac{x_{AH}}{x}\right)} + \bar{h}(x_{AH}),
\end{equation}
with $\ln x_{AH} = s_\infty$ and $x\leq x_{AH}$. Hence, the scalar field maintains finite and continuous at the apparent horizon and may be further extended to infinity.

%
\begin{table}
  \caption{\label{table} Summary of important points in the phase space. }
  \begin{ruledtabular}
  \begin{tabular}{ccccc}
  Points & Coordinates $(y,\zeta,\gamma)$ & Description & Stability & Eigenvalues \\ \hline
  $\mathcal{O}$ & $(1,0,0)$ & initial starting point $s=-\infty\ (u<0, r=0)$ & saddle & (\ref{eigen_O}) \\
  $\mathcal{P}_1$ & $(\frac{1}{2},\frac{2}{3+\kappa},1)$
  & critical ending point $s=+\infty\ (u=0, r>0)$ & saddle & (\ref{eigen_P1}) \\
  $\mathcal{P}_{-1}$ & $(\frac{1}{2},\frac{2}{3-\kappa},-1)$
  & critical ending point $s=+\infty$ & saddle & (\ref{eigen_P-1}) \\
  $\mathcal{P}_0(y)$ & $(y\neq 0,1,-\kappa)$
  & singular ending points $s=+\infty$, $\kappa\in (0,1)$ & attracting & (\ref{eigen_P0}) \\
  $\mathcal{P}_s(y)$ & $(y\neq 0,\frac{1}{2},\frac{1-y}{y\kappa^3})$
  & singular connecting points $s=s_*$, $\kappa\in (0,1)$ &  & (\ref{eigen_Ps}) \\
  $\mathcal{P}_{sc}$ & $(\frac{1}{1+\kappa^2},\frac{1}{2},\frac{1}{\kappa})$
  & singular connecting points $s=s_*$, $\kappa\in (0,1)$ &  & (\ref{eigen_Ps}) \\
  $\mathcal{P}_{AH}$ & $(0,\frac{1}{2},\pm\infty)$
  & apparent horizon $s=s_\infty$ & attracting & \\
  \end{tabular}
  \end{ruledtabular}
\end{table}

\section{Global solutions} \label{sec:global}

We move on to constructing global solutions via numerical integration and verifying our previous analysis on local behaviors. To systematically create numerical curves from the initial (stationary) point $\mathcal{O}$, our strategy is to use the truncated series solution (\ref{seriesL}) or (\ref{seriesT}) to generate initial values near $\mathcal{O}$ (e.g., taking $s=-10$ and $c_3=1,2,\cdots$), and then integrate both backward and forward with the standard fourth-fifth order Runge-Kutta method. This simplifies numerical procedures and fulfills the conditions (\ref{IV}) and (\ref{IV1}) to a satisfactory accuracy. Moreover, upon nearing the singular curve $\mathcal{P}_s(y)$ or the plane $y=0$, it is also desirable that step sizes of integration be kept small so as to produce more accurate plotting. Similar procedures also apply to integral curves from other critical points.

As we have seen in local studies, the nature of the critical and singular points relies heavily on the value of the parameter $\kappa$. Our numerical results split into two cases.

\subsection{Case $\kappa > 1$} \label{sec:case1}

In our first set of plots with $\kappa=1.2$, solutions of interest are limited to $0<\zeta<1/2$ with $\mathcal{P}_1$ involved. Since $\mathcal{P}_1$ is a saddle point, the 1-dimensional unstable manifold (red curve) can draw away incoming integral curves (green and cyan) from $\mathcal{O}$ on either side of the 2-dimensional stable manifold (cf. Sec. \ref{sec:critpt}). This naturally classifies the solutions into three types. For solutions heading toward the plane $y=0$ (green curves), they have $c_T>\kappa^2/3$ and correspond to $V_0=1$ by (\ref{V0_cT}). Numerical integration also indicates that they quickly exceed the maximal bound of the programme within a finite $s$ while ascending in the direction of $\mathcal{P}_{AH}$ (cf. (\ref{gammaAH})). This first type represents black hole formation. On the other side of the stable manifold, solutions moving away from $y=0$ (cyan curves) have $V_0=-1$ and are deemed unphysical in the sense that they severely violate the positive mass condition $y\leq 1$ at a large $s$ (in fact, the mass function can go arbitrarily negative). The intermediate between the two types is an exceptional solution (blue curve) terminating at $\mathcal{P}_1$ as $s\rightarrow+\infty$. It also lies in the massless phase subspace determined by (\ref{eqaml}), hence corresponding to the unique interior solution by Christodoulou \cite{Christo94}. According to \cite{Brady95}, the associated spacetime develops a central singularity at $u=0$ that is not visible to observers at a finite radius.

Unlike the massless evolution, black hole formation completely takes over under positive exponential potentials. This ``regularizing'' effect appears in line with cosmic censorship. Nevertheless, negative exponential potentials seem to have an opposite effect by rendering the spacetime more pathological. This comparison will be further discussed in later sections. For other values of $\kappa>1$, the qualitative behavior we have described remains the same.

\subsection{Case $0<\kappa < 1$} \label{case2}

\begin{figure}[htbp]
  \centering
  \mbox{
  \subfigure{\includegraphics[width=0.48\textwidth]{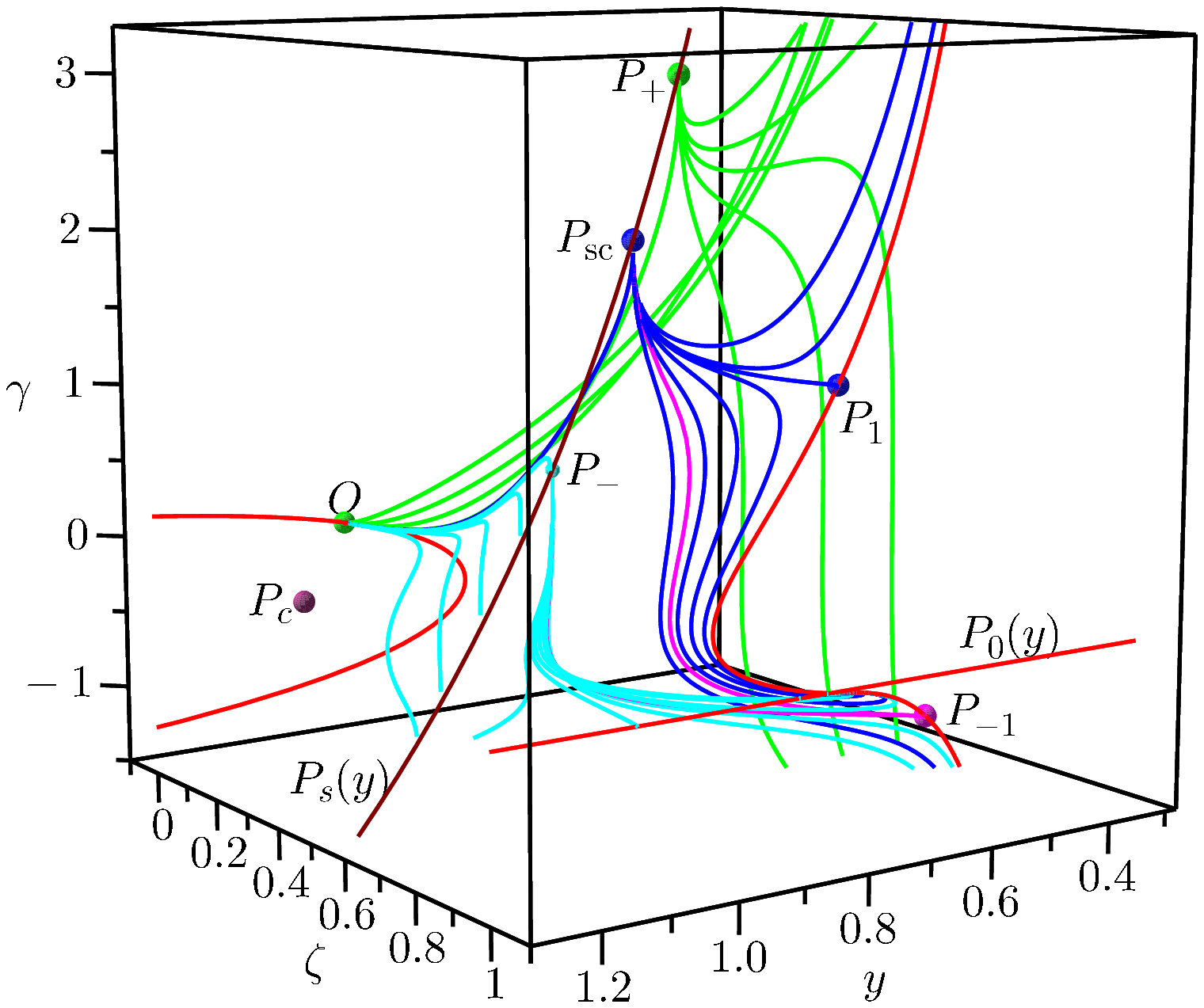}}
  \subfigure{\includegraphics[width=0.48\textwidth]{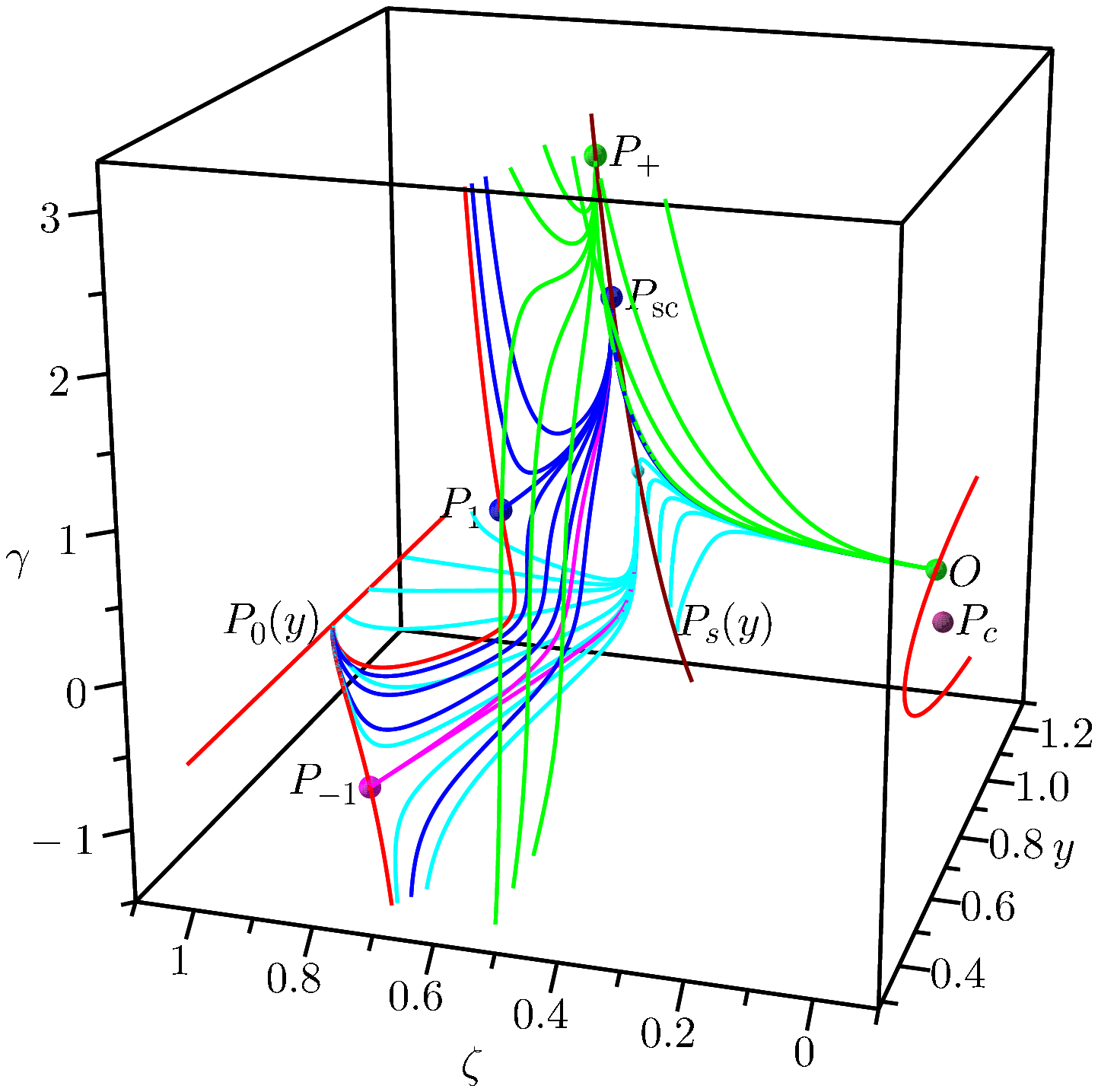}}
  }
  \caption{Numerical integration with $\kappa=0.5$ (roughly front and back views). Integral curves from $\mathcal{O}$ may reach the singular curve $\mathcal{P}_s(y)$ (maroon). Particularly, the blue ($V_0=0$) curve attaining $\mathcal{P}_{sc}$ is the unique massless interior solution interpolating between green ($V_0=1$) and cyan ($V_0=-1$) curves. In the region $\zeta>1/2$, curves drawn from $\mathcal{P}_{\pm}$ and $\mathcal{P}_{sc}$ are continuations of the interior solutions. The two purple curves terminating at $\mathcal{P}_{-1}$ separate naked singularity solutions and black hole solutions. The red curve passing through $\mathcal{P}_1$ ($\mathcal{P}_{-1}$) with one end reaching $\mathcal{P}_0(y)$ represents the unstable manifold of $\mathcal{P}_1$ ($\mathcal{P}_{-1}$) in the massless phase subspace (\ref{eqaml}).
  }
  \label{fig2}
\end{figure}

\begin{figure}[htbp]
  \centering
  \mbox{
  \subfigure{\includegraphics[width=0.48\textwidth]{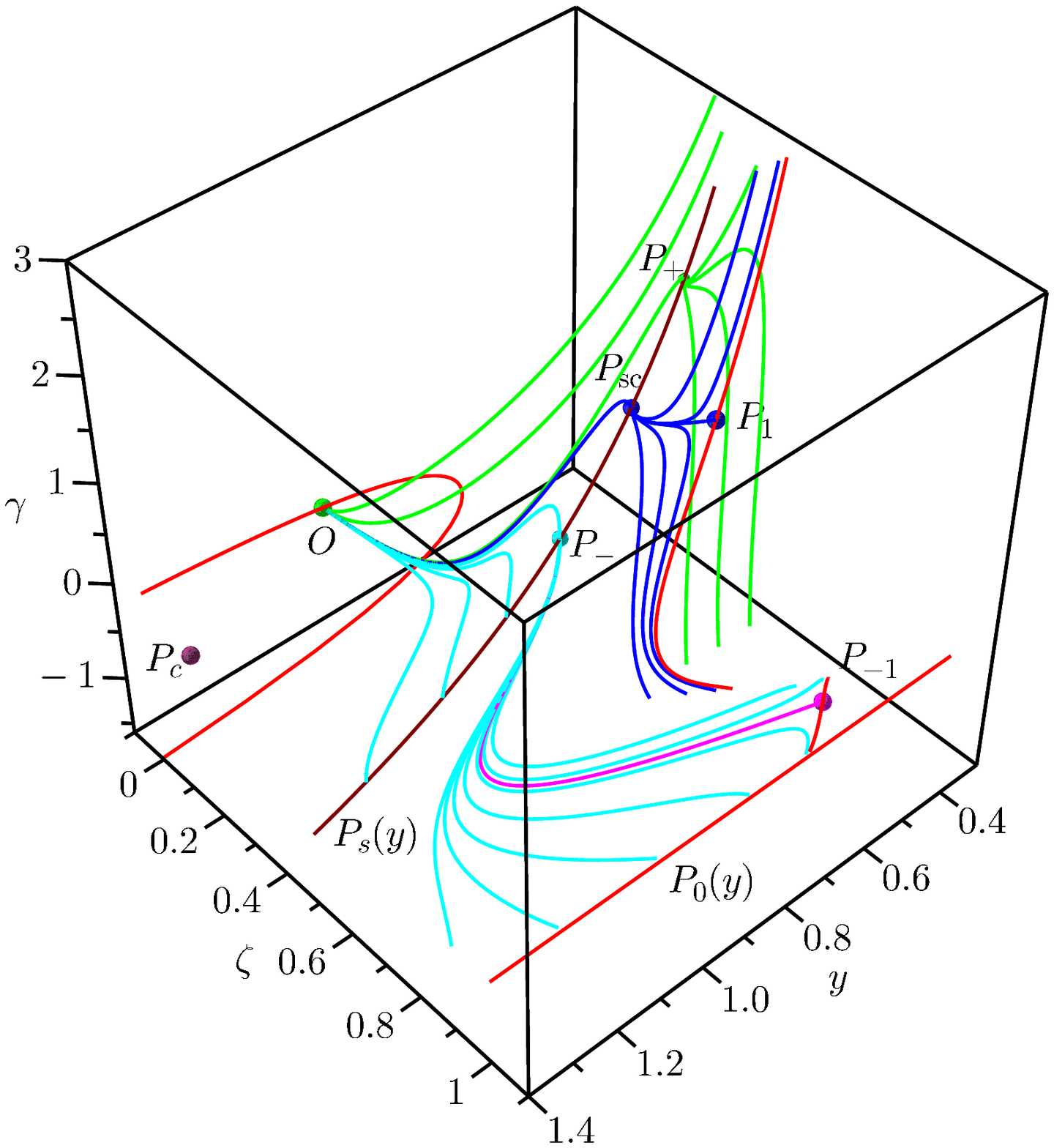}}
  \subfigure{\includegraphics[width=0.48\textwidth]{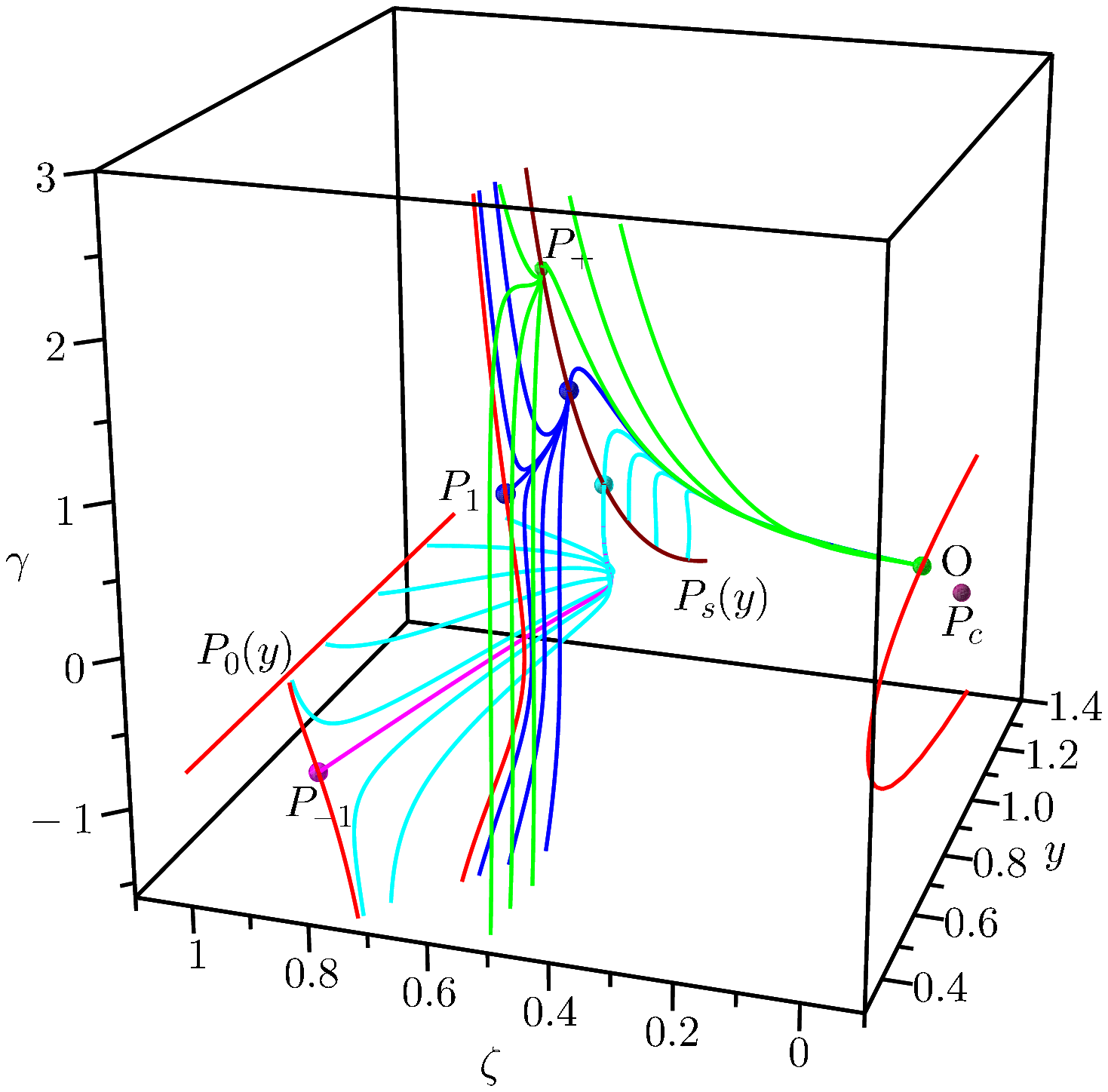}}
  }
  \caption{Numerical integration with $\kappa=0.7$ (roughly upper front and back views). Behavior of the green ($V_0=1$) and cyan ($V_0=-1$) curves are similar to those in Fig. \ref{fig2}. However, for massless solutions, the blue curves and the unstable manifold of $\mathcal{P}_1$ (red curve) no longer extend to the singular line $\mathcal{P}_0(y)$, indicating that naked singularities cease to occur.}
  \label{fig3}
\end{figure}

This case is much more involved as integral curves can pass through the plane $\zeta=1/2$. In the region $0<\zeta<1/2$, an interior solution curve shown in Fig. \ref{fig2} and Fig. \ref{fig3} can either move toward the apparent horizon $y=0$ or run into the singular curve $\mathcal{P}_s(y)$ in a finite $s$. In Sec. \ref{sec:singlpt}, we point out that once attaining a point on $\mathcal{P}_s(y)$, an interior solution may branch into a one-parameter family of exterior solutions. This is confirmed in Fig. \ref{fig2} and Fig. \ref{fig3}, where we have picked out three representative locations on $\mathcal{P}_s(y)$---$\mathcal{P}_{sc}$ for massless solutions and $\mathcal{P}_{\pm}$ for $V_0=\pm 1$---from which continuations are plotted. Amongst these extended curves, naked singularity solutions correspond to those approaching $\mathcal{P}_0(y)$ as $s\rightarrow +\infty$. Clearly in Fig. \ref{fig2} with $\kappa=0.5$, such curves can emerge from $\mathcal{P}_{sc}$ (also $\mathcal{P}_-$, but not from $\mathcal{P}_+$). However, when $\kappa=0.7$ as in Fig. \ref{fig3}, they cease to show up in massless solutions altogether (blue curves). This numerical result agrees with Christodoulou's rigorous argument that, in massless scalar fields, naked singularities only exist when $\kappa^2<1/3$. To understand this phenomenon more intuitively from our plots, we note that the unstable manifold $\mathcal{M}_1$ (red curve, restricted on the surface (\ref{eqaml})) of the saddle point $\mathcal{P}_1$ appears to be a good indicator since it may dictate asymptotical behavior of the integral curves in its vicinity. More specifically in Fig. \ref{fig2}, one arm of $\mathcal{M}_1$ extends to $\mathcal{P}_0(y)$, thereby pulling solution curves nearby to the same endpoint. While in Fig. \ref{fig3}, both ends of $\mathcal{M}_1$ lead away from $\mathcal{P}_0(y)$, and are followed by other integral curves. Moreover, according to numerics, the critical case of $\kappa=1/\sqrt{3}$ has the lower segment of $\mathcal{M}_1$ falling exactly on the point $\mathcal{P}_{-1}$ \cite{Christo94}. No naked singularity arises in this critical case.

The situation, however, differs greatly for solutions with negative exponential potentials (cyan curves). Starting from a singular connecting point lower than $\mathcal{P}_{sc}$, an exterior solution may evade the influence of $\mathcal{P}_1$ and still reach out to $\mathcal{P}_0(y)$ regardless of $k^2<1/3$ or $1/3\leq k^2<1$. The only caveat here is that solutions of such may locally violate the positive mass condition $y\leq 1$ and may even have negative total masses (see Fig. \ref{fig3}). A similar situation was also encountered in \cite{Hertog04a}. Besides naked singularities, other outcomes are also presented. In both Fig. \ref{fig2} and Fig. \ref{fig3}, by tuning the second parameter, one can observe a transition between naked singularity and black hole solutions. The boundary separating these two extremes is demarcated by critical solutions terminating at $\mathcal{P}_{-1}$ (purple curves; more generally, the stable manifold of $\mathcal{P}_{-1}$). The resulting spacetime can be thought as sitting on the threshold of containing a naked singularity or a black hole, but just avoiding both. Similar critical behavior is also found in massless solutions (blue curves near $\mathcal{P}_1$ and $\mathcal{P}_{-1}$ in Fig. \ref{fig2}). As an aside, see \cite{Zhang15a,Zhang15b} for more examples on naked singularity/black hole transitions with non-trivial scalar potentials.

As regards solutions with $V_0=1$, black hole formation appears to be the only possible outcome as all green curves are heading to $\mathcal{P}_{AH}$ instead of $\mathcal{P}_0(y)$. This observation is further supported by numerical tests on other locations on $\mathcal{P}_s(y)$ above $\mathcal{P}_{sc}$. Therefore, we speculate that the naked singularity solutions by Christodoulou defy extensions to positive exponential potentials.

\section{Spacetime Structure} \label{sec:struc}

In this section, we examine various aspects of the metric (\ref{metricx}) with inputs from the solutions we have computed. To begin with, we show that for solutions terminating at $\mathcal{P}_0(y)$, the corresponding spacetime indeed contains a curvature singularity at the point $u=0$, $r=0$ that is visible to distant observers. To see this, one can calculate the Ricci scalar as
\begin{eqnarray}
 R &=& - 2 T = 2 \big(\nabla_\alpha \phi \nabla^\alpha \phi + 4V(\phi) \big) \\
 &=& 2 \left[ -\frac{2}{g} \phi_{,u}\phi_{,r} + \frac{\bar g}{g}(\phi_{,r})^2 + 4V_0 \exp\!\left(\frac{2\phi}{\kappa}\right)\right] \\
 &=& \frac{2\bar g}{u^2 x^2 g} \left[\gamma^2 - \frac{2x}{\bar g} (\gamma^2+\kappa\gamma)\right] + \frac{8V_0 w}{u^2x^2} \\
 &=& \frac{2y}{u^2 x^2} \left[\gamma^2 - \frac{\zeta}{1-\zeta}(\gamma^2+\kappa\gamma)\right] + \frac{8V_0 w}{u^2x^2}. \label{Ricci}
\end{eqnarray}
Then by applying the asymptotic series (\ref{P0_zgamma}) and the limit $y\rightarrow y_0$ as $s\rightarrow +\infty$ (equivalently $u\rightarrow 0-$), one arrives at
\begin{equation}
 R = \frac{1}{r^2} \left[2(y_0-1) \left(\frac{1}{\kappa^2} - 2\right) - 4y_0\kappa^2\right],
\end{equation}
with $y_0=1/(1+\kappa^2)$ for the massless solutions. Clearly, this scalar invariant blows up at $r=0$, but remains finite elsewhere on the future light cone $u=0$ of the center (the Cauchy horizon). Therefore, the spacetime may be further extended beyond $u=0$, at which the original self-similar coordinate fails.

Now it remains to establish that the central curvature singularity is indeed visible, i.e., that the outgoing null geodesics $u=0$ can escape from $r=0$. Following the method in \cite{Christo94} (see also \cite{Bedjaoui10}), we investigate incoming radial null geodesics, assuming $\zeta>1/2$, and show that they can approach a finite $r$ as $u\rightarrow 0$. In terms of the variables $t=-\ln(-u)$ and $s=\ln(-r/u)$, the null geodesic equation
\begin{equation}
 \frac{\rmd r}{\rmd u} = -\frac{\bar{g}}{2}
\end{equation}
can be rewritten as
\begin{equation}
 \frac{\rmd s}{\rmd t} = -\frac{1}{\zeta(s)} + 2 > 0.
\end{equation}
Hence, along a null ray originated at some $(t_0, s_0>s_*)$, we have
\begin{equation}
 t - t_0 = \int^s_{s_0} \frac{\zeta}{2\zeta-1} \rmd s'.
\end{equation}
The asymptotic series (\ref{P0_zgamma}) implies that
\begin{equation}
 \left(\frac{\zeta}{2\zeta-1} - 1\right) \rme^{(1-\kappa^2)s} \rightarrow -c_2,
 \qquad s\rightarrow +\infty.
\end{equation}
Therefore, the quantity
\begin{equation}
 \ln r = s - t = s_0 - t_0 - \int^s_{s_0} \left(\frac{\zeta}{2\zeta-1} - 1\right) \rmd s'
\end{equation}
converges to a finite limit as $t\rightarrow +\infty$ since the integral above is bounded. Also because $c_2<0$ from numerical results, we conclude that along an incoming null ray near $u=0$, the radius $r$ decreases to a finite value as $u$ increases to $0$.

Using the limit $y\rightarrow y_0$ as $s\rightarrow+\infty$, we can further determine the mass function at $u=0$ as
\begin{equation}
 m = \frac{1-y_0}{2}\, r,
\end{equation}
with $y_0=1/(1+\kappa^2)$ for the massless solutions and $y_0>1/(1+\kappa^2)$ for $V_0=-1$. Thus the central singularity itself has vanishing mass. Similarly, we can write down the scalar field in the limit $u\rightarrow 0$ as
\begin{equation} \label{phi}
 \phi = \frac{\kappa}{2}\ln\!\left(\frac{1-(1+\kappa^2)y_0}{2V_0}\right) - \kappa \ln r,
 \qquad V_0=-1, \qquad y_0>1/(1+\kappa^2),
\end{equation}
and for the massless scalar fields,
\begin{equation}
 \phi = -\kappa \ln r,
\end{equation}
both of which diverge logarithmically at $r=0$. From the equation (\ref{phi}), the negative potential energy at $u=0$ obeys
\begin{equation}
 V \propto -\frac{1}{r^2},
\end{equation}
which is non-negligible near the center.

Lastly, we comment that the naked singularity solutions that we have considered are, by themselves, neither asymptotically flat nor AdS. Such unusual asymptotic is in fact typical for negative exponential potentials \cite{Poletti94,Chan95,Cai04}. Nevertheless, we should emphasize that since the past light cone of the singularity is given by $-r/u = \exp(s_*)$, the naked singularity formation is essentially a local event near the symmetry center and independent of the geometry far from it.

\section{Concluding remarks} \label{sec:conclusion}

In the class of self-similar spacetimes, we have studied the effect of exponential potentials on the naked singularity solutions that were first discovered by Christodoulou and numerically explored by Brady. For every fixed $\kappa$, the original 2-dimensional phase space is embedded into a 3-dimensional one, which allows more dynamical events to take place. Using numerical methods, we build a pictorial representation of Christoloulou's solutions and put them in a richer context with new solutions. In our 3-dimensional plots, two competing ``attractors''---one for naked singularities and one for black holes---play a major role as a time evolution may be drawn into either one of them. On one side of Christodoulou's solution manifold, a \emph{two}-parameter family of naked singularity solutions continues to exist under the negative potentials $V=-\exp(2\phi/\kappa)$ for $0<\kappa^2<1$. In contrast, on the side with ``regularizing'' positive exponential potentials, all numerical evolutions tested by us encounter apparent horizons, which suggests that the naked singularity solutions by Christodoulou may be unstable against perturbation of the positive potentials. A similar effect is also found in the regime $\kappa^2>1$ (see Sec. \ref{sec:case1}).

Regarding future work, it would be natural to investigate embedding of our solutions in higher dimensions \cite{Cai04,Mignemi89}. In a follow-up research \cite{An15}, we will show that such embedding leads to naked singularity solutions in higher dimensional vacuum, thereby putting our current studies in a fruitful perspective. To conclude, our results may shed light on the issue of cosmic censorship in string theory.


\appendix*

\section{Christodoulou's notation} \label{app:Christo}

One can translate Brady's notation that we have been using to Christodoulou's \cite{Christo94} through
\begin{equation}
 y^{-1}=\rme^{2\lambda}, \qquad z^{-1}=2(1-\beta), \qquad \gamma = \theta, \qquad \kappa = k.
\end{equation}
In terms of the latter, the system (\ref{ode1}-\ref{ode3}) reads
\begin{eqnarray}
 \frac{\rmd \lambda}{\rmd s} &=& -\frac{(\theta+k)(2\beta\theta-\theta+k)}{2(1-\beta)}, \label{odec1}\\
 \frac{\rmd \beta}{\rmd s} &=& 1-k^2 - \left[(\theta+k)^2+1-k^2\right]\beta,  \label{odec2}\\
 \frac{\rmd \theta}{\rmd s} &=& \frac{k}{\beta}(k\theta-1)+\left[(\theta+k)^2-(1+k^2)\right]\theta
 + \frac{1-\beta}{k\beta}
 \left[ 1+k^2 + \frac{\beta}{1-\beta}(\theta+k)^2 - \rme^{2\lambda}\right]\!\!. \label{odec3}
\end{eqnarray}
Also the constraint (\ref{eqaml}) for massless scalar fields becomes
\begin{equation} \label{eqamlc}
 \rme^{2\lambda} = 1+k^2 + \frac{\beta}{1-\beta}(\theta+k)^2.
\end{equation}
Using this equation, we can remove $\lambda$ in (\ref{odec3}) and obtain a reduced system
\begin{eqnarray}
 \frac{\rmd \beta}{\rmd s} &=& 1-k^2 - \left[(\theta+k)^2+1-k^2\right]\beta, \\
 \frac{\rmd \theta}{\rmd s} &=& \frac{k}{\beta}(k\theta-1)+\left[(\theta+k)^2-(1+k^2)\right] \theta,
\end{eqnarray}
which recover the equations (0.27a,b) in \cite{Christo94}. It is straightforward to verify that if the above system holds for $\beta$ and $\theta$, $\lambda$ as determined by (\ref{eqamlc}) automatically satisfies (\ref{odec1}).

\begin{acknowledgments}
 We extend our gratitude to Daniel Finley, Hong L\"u, Sijie Gao, Christopher Pope, Hisaaki Shinkai, and Tong-Jie Zhang for helpful discussions and kind assistance. We also thank Luis Lehner for valuable suggestions and comments on an earlier version of the manuscript. The project was supported by China Postdoctoral Science Foundation and in part by the NSFC grants 11175269, 11475024 and 11235003. XA also acknowledges AIP Fund from Rutgers University.
\end{acknowledgments}


\end{document}